\let\myover=\over   
\documentclass[12pt,a4paper]{article}
\usepackage{epsfig}
\def\be{\begin{equation}}
\def\ee{\end{equation}}

\def\l{\left(}
\def\r{\right)}

\def\be{\begin{equation}}
\def\ee{\end{equation}}

\def\half{\frac{1}{2}}

\newcommand{\bg}{\begin{gather}}
\newcommand{\eg}{\end{gather}}

\newcommand{\lesssim}{\le}

\begin{document}
\let\over=\myover  
\def\half{{1 \over 2}}

\begin{titlepage}
\thispagestyle{empty}

\vspace*{3cm}
\begin{center}
\Large \bf Stoponium Search at Photon Linear Collider
\end{center}

\vspace*{2cm}

\begin{center}
\large 
D.S.~Gorbunov$^*$
\ \ and \
V.A.~Ilyin$^\dagger$
\end{center}

\begin{center}
{\small\em
$^*$Institute for Nuclear Research, Moscow, 117312, Russia} \\
{\small\em
$^\dagger$Institute of Nuclear Physics, Moscow St. Univ., Moscow, 119899, Russia
}       
\end{center} 
       
\begin{quotation}
\vspace*{1.5cm}
        
\vskip 2cm  
\begin{center}
{\sf Abstract}
\end{center}
 
In some supersymmetric extensions of the Standard Model fairly light
superpartner of t-quark is predicted, which may form bound states ({\it
stoponiums}) under certain conditions. We study prospects of search for
stoponium at the future Photon Linear Collider. It is found that  this machine
could be the best machine for discovery and study of these resonances at some
scenarios of supersymmetric extension of the Standard Model. In particular, if
the $hh$ decay channel is dominant stoponium could be observed at the beginning
of PLC run with collision energy tuned at the stoponium mass. If this
channel is kinematically closed stoponium could be 
discovered in $gg$, $\gamma\gamma$ 
and $ZZ$ decay channels but higher statistics are needed. 
Effects of the
stoponium-Higgs mixing and degeneracy are briefly discussed.

\end{quotation}

\end{titlepage}

\vspace{0.5cm}
\noindent
{\large \bf Introduction}
\vspace{0.2cm}

The broken supersymmetry is favorite among the different extensions of  the
Standard Model. It can happen that superpartners of top-quarks ({\it stops},
$\tilde{t}$) are long-living enough to compose (colorless) bound states, {\it
stoponiums}, denoted as $S$ in what follows. In this scenario experimental
study of the corresponding resonances could provide precise value of stop mass
and stoponium partial widths, consequently yielding precise values of various
parameters of SUSY Lagrangian. Then, if the difference between stop and LSP
masses is very small, the search for stop evidence in collisions at high energy
could be problematic. Observation of stoponium bound states will be the
signature of such models confirming the existence of stop. 

There are theoretical motivations for stop to be fairly light. First one
appeals to the renormalization group  behavior of soft mass terms. Indeed,
gauge couplings raise while Yukawa couplings reduce these terms when energy
scale evolves down, with Yukawa contributions being very large for stop. The
next motivation concerns left-right mixing in squark sector, which is
proportional to Yukawa coupling and decreases the mass of the lightest stop.
Therefore, light stop may appear in different SUSY models ( see, e.g., Refs.
\cite{light-stop} for examples in the frameworks of supergravity and gauge
mediation).  Experimental bound on stoponium mass comes from searches for stop
at LEP2 and TEVATRON. The concrete number depends on the MSSM
spectrum~\cite{stop-exp}: lower bound is about 90 GeV for sneutrino masses
larger than 45-50 GeV or for neutralino masses larger than 50 GeV (ALEPH),
while CDF excludes stop mass up to 130 GeV for smaller sneutrino masses. The
limitation is weaker if stop and neutralino masses are degenerate, it is about
60 GeV (ALEPH).

Stoponium was studied in Refs.~\cite{stoponium,drees} in detail, in particular
its effective couplings and partial widths were calculated, and probability to
be discovered at LHC were estimated. In Ref.~\cite{drees} it was  briefly
mentioned also the possibility to observe stoponiums in photon collisions,
however, without analyzing this phenomenology. 

Now, when main characteristics of future Photon Linear Colliders
(PLC)\footnote{The possibility to get real photon beams by means of Compton
backscattering of laser photons on high energy electrons from the basic
electron beams was proposed a long time ago~\cite{backscattering}.} are under
technical discussion (see, e.g., Ref.~\cite{PLC}) one should understand clearly
signatures of  stop bound states in this type of high energy collisions. This
is of special interest due to  debates on PLC advantages in comparison with
further increasing of the collision energy for basic electron-positron mode of
the linear collider when the run at $\sqrt{s_{e^+e^-}}=500$ GeV will be
completed (see, e.g., Ref.~\cite{telnov} and references therein). One can
easily perceive that $e^+e^-$ colliders (LEP2 or linear electron-positron
colliders) have no good chances to observe stoponiums. Indeed, high powers of
the coupling constants, $\alpha^2\alpha_s^5$, emerge in the squared matrix
elements: $\alpha^2$ arises from two electroweak vertices, and $\alpha_s^5$
comes from squared derivative of the stoponium wave function (scalar stoponium
can be created there only in P-wave  by propagation of vector particle, photon
or $Z$). At the same time two powers of $\alpha_s$ are eliminated in the case
of $\gamma\gamma$ collisions since stoponium can be produced in S-wave. So, one
can expect much higher rate of the stoponium production at PLC.

In hadron collisions the stoponium production is also available in S-wave
through the gluon fusion. So, effective $ggS$ vertex includes $\alpha_s^{5/2}$
and one can anticipate large stoponium cross sections as well. However, main
decay channel $gg$ is too dirty due to huge QCD $2jets$ background. Then, as
it was found in Ref.~\cite{drees}, the most promising decay channel at LHC is
$\gamma\gamma$, but in order to discover stoponiums  one year of LHC operating
at high luminosity is needed, or even more (depending on SUSY scenario). 

It is shown in this letter that Photon Linear Collider will be the best machine
to discover and study stoponiums. For estimates we use basic parameters of the
first stage of TESLA project: $\sqrt{s_{ee}}\le 500$ GeV and integrated
luminosity 500 fb$^{-1}$~\cite{ee500} for basic electron-positron mode. It is
well known that high energy photon beam at the PLC conversion point will have a
rather wide spectrum \cite{backscattering}. However, there are various ideas
under discussion how to get more or less monochromatic photon beams. So, for
estimates one can take the photon-photon collision energy spread within the
interval of 15\% below the maximal energy $E^{max}_{\gamma\gamma}\sim 0.8
\sqrt{s_{ee}}$ \cite{PLC}. We set effective integrated luminosity of {\it the
first} PLC equal to 100 fb$^{-1}$ for this 15\% window, although one can judge
this figure as too pessimistic --- in Ref.~\cite{PLC} this characteristic is
argued to be at the same level as for basic $e^+e^-$ mode.

We consider stoponium mass range $M_S=200-400$~GeV, which could be surely 
probed by {\it the first} PLC. Of course tuning the $\gamma\gamma$ collision
energy at the stoponium mass point is necessary. It is worth to note that the
same interval is not an exceptional case for SUSY models with stoponiums as a
quasistationary state, as we discuss briefly in the next section.

\vspace{0.5cm}
\noindent
{\large \bf Stop bound states}
\vspace{0.2cm}

It is clear that gluons try to bind two stops as well as ordinary quarks. The
corresponding bound state can be described as a quasistationary system with
energy levels $E_n$ ($<0$) and masses $M_n=2m_{\tilde{t}}+E_n$ similarly to
quarkonium. For stoponium mass $M_S=200-600$~GeV the binding energies $E_n$ are
of order 1 GeV~\cite{ng}. This treatment is valid if the formation process
(time scale $\sim |E_n|^{-1}$) is faster than destroying one.  

Among destroying mechanisms the obvious ones are the stop decays\footnote{We
suppose R-parity to be conserved, as favored by the absence of rapid proton
decay and lepton flavor violating processes.}: $\tilde t\to t+$LSP,
$b+chargino$ and $c+neutralino$. At first, let us consider the third decay. It
proceeds only through loop diagrams, that is motivated by the absence of FCNC
(thus, leading to universality-like soft terms). So, partial width is highly
reduced by a factor of $\sim 10^{-7}$ in comparison with the first two
tree-level decay processes~\cite{c-neutralino}. The rates of latter decays
depend on the parameters of the model. As an example, in the framework of
gravity mediation, where LSP is neutralino, these decays proceed at the tree
level and the corresponding partial widths are of order ${\cal O}(\alpha
m_{\tilde t})$. In the framework of models with gauge mediated supersymmetry
breaking \cite{revGMM}, where LSP is gravitino, the first process is strongly
suppressed by supersymmetry breaking scale, but remaining one has the same
partial width as in gravity mediation. Hence, the possibility of existence of
stoponium is a subject of special study in each concrete model. For instance,
in models with the lightest chargino being mostly wino and the lightest stop
being mostly right stop (i.e., $m_{t_L}>m_{t_R}$), decay into chargino is
damped and stoponium could exist if $m_{\tilde t}-m_{LSP}<m_t$, i.e., when the
first decay channel is kinematically forbidden. One can state that SUSY
scenario, where tree-level decay channels, $\tilde t\to t+$LSP and $\tilde t\to
b+chargino$, are somehow suppressed and, therefore, stop decay can not destroy
the stoponium formation, is not an exceptional case.

Next destroying mechanism is related to the stop annihilation. Here two gluon
channel is always open with partial width  about 1 MeV. Generally the gluon
channel is dominant. However, for the certain choice of model parameters,
partial width into two lightest Higgs bosons, $S\to hh$, can be larger,
increasing the stoponium total width by a factor of $\sim 5-10$. In
Ref.~\cite{drees} these figures were analyzed and found that quasistationary
description is valid for $M_S<600$~GeV in models with forbidden stop tree-level
decays and neutralino being mostly bino. The worst case is a model with
chargino and neutralino states are both higgsino-like. Here the stoponium total
width increases rapidly with $M_S$ and quasistationary treatment fails for
$M_S>300$~GeV. 

To be concrete let us describe one of the models studied in Ref.~\cite{drees}, 
where stoponium exists up to $M_S<600$~GeV. This is MSSM with $\mu=-300$~GeV,
$m_{t_L}=400$~GeV, $m_{t_R}=300$~GeV and mass of the lightest stop (so, $M_S$)
varied with trilinear soft term $A_t$. The $hh$ decay channel  dominates at
$2m_h\lesssim M_S\lesssim 2m_h+150$~GeV. Gluonic decay mode prevails in other
cases, excluding the vicinity of the degenerate point with Higgs boson mass
$M_S=M_H$  where $b\bar b$ or $t\bar t$ channels becomes principal because of
s-channel Higgs propogator enhancement. In scenarios with
$A_t=2m_{t_L}=2m_{t_R}$ and $\mu=500$~GeV decay into $hh$ final state
dominates  for wider range, $2m_h\lesssim M_S\lesssim 600$~GeV. 

\vspace{0.5cm}
\noindent
{\large \bf Stoponium in $\gamma\gamma$ collisions}
\vspace{0.2cm}

The main effect associated with stoponium would be a direct resonance
production. The corresponding monochromatic cross section can be written in the
Breit-Wigner form if collision energy is close to the resonance peak 
$$
\sigma_{\gamma\gamma\to S\to f}(\hat{s})= 8\pi \cdot
\frac{\Gamma_{\gamma\gamma}\,\Gamma_f}{\l \hat{s}-M_S^2\r^2+ \Gamma_{\rm
tot}^2M_S^2} \;, 
$$

\noindent
where $\Gamma_f$ is the stoponium partial width for the decay into state $f$
and $\Gamma_{\rm tot}$ is stoponium total width.  At the resonance point this
cross section reads $\sigma^{res} = {8\pi\over M_S^2}\; Br_{\gamma\gamma}\;
Br_f,$ where $Br_f={\Gamma_f/\Gamma_{\rm tot}}$ is the corresponding branching
fraction. The observable cross section can be estimated as
$\sigma^{res}\cdot2\Gamma_{\rm tot}/(0.15 E^{max}_{\gamma\gamma})$ with
$E^{max}_{\gamma\gamma}\sim M_S$. Photon beams are planned to be highly
polarized. Hence, as stoponium is a scalar the production cross section will be
enhanced by factor two if initial photons have opposite helicities.  Hereafter
we include this factor 2 in our estimates of signal rates. Finally, we
parameterize the stoponium cross sections as follows
\begin{equation}
\sigma_f \;\approx\; 130\mbox{fb} \cdot 
   \left(\frac{Br_{\gamma\gamma}}{4\cdot 10^{-3}}\right) \cdot
   Br_f \cdot
   \left(\frac{\Gamma_{tot}}{1\mbox{MeV}}\right) \cdot
   \left(\frac{200\mbox{GeV}}{M_S}\right)^3 .
\label{gamgam}
\end{equation}
where, in addition, one should take into account that squared stoponium wave
function at the origin, attending in $\Gamma_{tot}$,  scales as a square root
of its mass~\cite{ng}.  

As it has been stressed above one can discuss two main variants of the SUSY
models, one with dominant $gg$ decay mode and another with stoponium total
width being saturated by $hh$ mode. Let us make qualitative {\it
signal/background} estimates for different decay channels within these two
variants. The signal significance can be evaluated by ratio $N_S/\sqrt{N_B}$
because one deals with resonance and background rate in the signal bin can be
fixed as average cross section in neighboring bins (here $N_{S,B}$ are numbers
of signal and background events). We used the results for stoponium width and
branching ratios calculated in Ref.~\cite{drees} with some corrections. Namely,
comparing our formulas with formulas in Ref.~\cite{drees} we conclude that
stoponium partial widths into two gluons and into two photons were
overestimated in Ref.~\cite{drees} because of extra factor 2 in Eqs.~(A.1) and
(A.2). In particular, it means that the signal cross sections were
overestimated there. In the first scenario for $gg$ and $\gamma\gamma$ modes
the correct cross sections are smaller by factor 2. In the second scenario for
$gg$ and $\gamma\gamma$ modes this factor is 4, while for other modes it is 2. 

Cross sections for various tree-level background processes were evaluated with
the help of CompHEP package \cite{CompHEP}. 

\begin{figure}[htb]
\begin{center}
{\epsfig{file=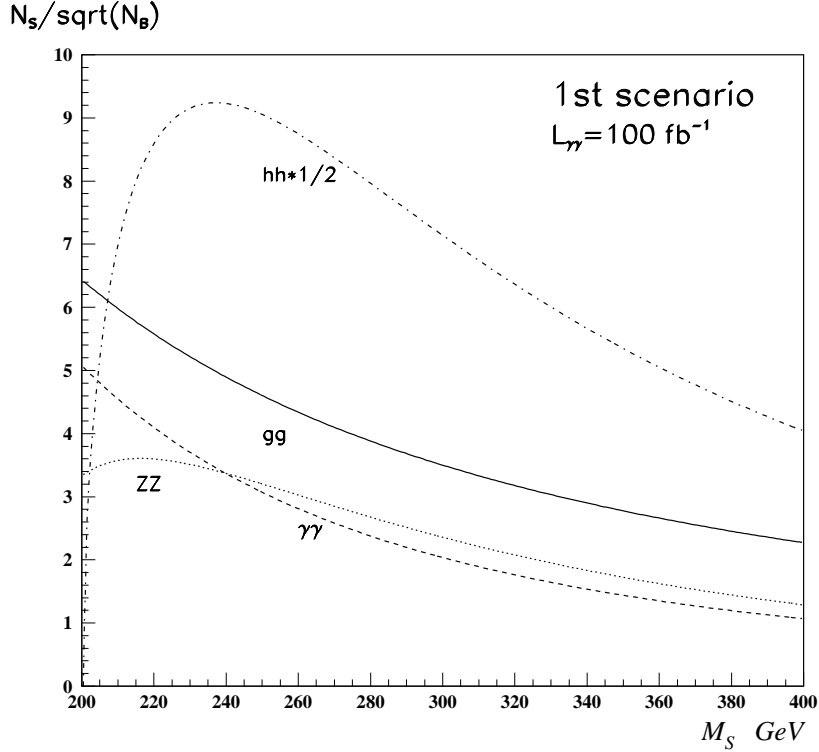,width=12cm}}
\end{center}
\caption{Signal significance of stoponium events in various channels for
the first scenario; mass of the lightest higgs boson is taken equal to
100~GeV.}
\label{gg-fig}
\end{figure}

\vspace{0.2cm}    
\noindent
{\bf 1.} 
In the first scenario stoponium total width $\approx 1.3$~MeV at
$M_S=200$~GeV and photon-photon
branching is ${\rm Br}_{\gamma\gamma}\approx 3.4\cdot 10^{-3}$. 
By making use of
Eq.(\ref{gamgam}) one obtains the signal rate at the level of 140 fb for
$M_S=200$ GeV. So, more than ten thousand stoponiums will be produced.
Background is two jet production, where subprocess $\gamma\gamma\to q\bar q$
gives main contribution with very large cross section, $\sim 50$ pb
(if cuts on the jet angle of 5$^\circ$ are applied). Since jet energy
resolution is of order ${\cal O}(10)$ GeV, background events should be counted
within whole 15\% photon-photon energy window. One can estimate background rate
directly as the average cross section in this window, $\sim 50$ pb. Hence
dominant $gg$ channel is fairly dirty --- the {\it signal/background} ratio is 
$\sim 1/350$. Nevertheless, for statistics 100 fb$^{-1}$ signal significance is
not too small, about $6\div 3$ for $M_S=200\div300$~GeV, see Fig.~\ref{gg-fig},
where we present signal significance for stoponium events in various channels. 
Here the $2jets$ background was calculated as a direct two-quark production,
thus with the rate being dependent on the collision energy.

For two photon channel the background process, $\gamma\gamma\to\gamma\gamma$,
proceeds through one-loop diagrams, so the corresponding cross section is
small, about 10 fb~\cite{JT1}. One should note that the  photon-photon
invariant mass bin can be taken equal to $2\mbox{GeV}\cdot
\sqrt{M_S/100\mbox{GeV}}$ for CMS-like crystal electromagnetic calorimeter
\cite{CMS}. Thus, for $M_{\gamma\gamma}=200\pm 1.4$ GeV  window the background
rate can be estimated as $\sim 1$ fb. For $M_S=200\div 300$~GeV the signal rate
is $0.5\div 0.2$ fb, providing the signal significance about $5\div2$.

Some other decay channels within the first scenario should be discussed. First
note, that $WW$ final state has no chance for the detection of stoponiums due
to huge SM background, $\sigma^{tot}_{\gamma\gamma\to WW}\sim 60$ pb at
$\sqrt{s_{\gamma\gamma}}=200$ GeV. More promising are decay channels with
background processes arising at higher orders of perturbation theory. For
instance, SM background to $\gamma Z $ and $ZZ$ final states comes from 1)
one-loop $\alpha^4$ processes $\gamma\gamma\to\gamma Z$ ($10-15$ fb~\cite{JT2})
and $\gamma\gamma\to ZZ$ ($\sim 50$ fb~\cite{ZZ}), and 2) from tree-level
$\alpha^3$ processes (e.g. $\gamma\gamma\to\gamma q\bar q$ for $S\to \gamma
Z\to\gamma +2jets$), with total cross section smaller than 1 fb within cuts on
final $\gamma$ and jets reasonably motivated by 2-body ($\gamma+Z$) kinematics
of the signal events. These figures should be considered as estimates of the
background rates in these channels for 15\% energy spread photon beams because
of the same level of the energy resolution for final jets.

As to the signal $\gamma Z$ rate one can get from Ref.~\cite{drees}  the
branching  ${\rm Br}_{\gamma Z} \sim 2\cdot10^{-3}$, so $\sigma_{\gamma Z}\sim
0.3$ fb already for $M_S=200$ GeV. It means very low level of the signal
significance, lower than 1 for statistics 100 fb$^{-1}$.

Natural level of $ZZ$ branching is about $4\cdot10^{-2}$ for stoponium masses
far from the threshold, 250-400 GeV, although in some points it could fall down
due to opening of new channels or degeneration of stoponium and Higgs masses.
This provides signal rate  $\sim 3.3$ fb for $M_S=250$ GeV and significance at
the level of 4.7  if one uses formula (\ref{gamgam}). However, the threshold
effect is significant still for this value of the stoponium mass, and these
figures should be improved to 2.25 fb for signal rate  and to 3 for
significance (see Fig.~\ref{gg-fig}).

The $hh$ decay channel, where $h$ is the lightest Higgs boson, is open if
$M_S>2m_h$. As current limit on $h$ mass is about 80-100 GeV this channel could
exist for $M_S>200$ GeV. If consider mass region far from the threshold (say
$M_S>230$ GeV for $m_h=100$ GeV) the $hh$ branching is about $2\cdot10^{-2}$ or
even higher. In this case the signal rate is about 2 fb or larger, 
if one uses formula (\ref{gamgam}), and if take into account the threshold
factor one gets signal cross section at the level of 1 fb or larger.

The background from direct double $hh$ production through one-loop diagrams can
be estimated by the cross section of this process in SM, $\sim 0.2$ fb
\cite{hh}. There are no reasons for very large additional contributions to this
process in supersymmetric models. Then, we found that direct electroweak
production of four b quarks ($\gamma\gamma\to bbbb$) together with contribution
from $cccc$ final state (assuming 10\% of $b/c$ misidentification) has the rate
smaller than 0.1 fb. Sum of these cross sections can be taken as the estimate
of the background rate for 15\% energy spread photon beams again, because
larger scale of the resolution bin for invariant mass of the stoponium decay
products. Therefore $hh$ decay mode will be the best channel for the
observation of stoponiums --- the signal significance is about 18 for
$M_S=230$~GeV. The signal rate is not so low, about hundred stoponium events
$S\to hh$ will be produced for statistics 100 fb$^{-1}$ and with $S/B$ ratio
about 3.

To resume the first scenario we conclude that, if stoponium decay into two
lightest Higgs bosons is allowed kinematically, this bound state could be
discovered after a few weeks of PLC operating. Otherwise, a year is necessary
in order to observe stoponiums in $gg$, $\gamma\gamma$, $ZZ$ channels. Then,
one can probe the stoponium effective couplings in all these channels if higher
statistics will be accumulated.

Note that LHC (at high luminosity operating stage) has good prospects to
observe light stoponium in $\gamma\gamma$ mode in this scenario~\cite{drees},
where signal significance is comparable with the significance of $hh$ mode at
PLC. Thus, these two colliders could be complementary in study of different
effective stoponium couplings, $S\gamma\gamma$ and $Shh$ correspondingly.

\begin{figure}[htb]
\begin{center}
{\epsfig{file=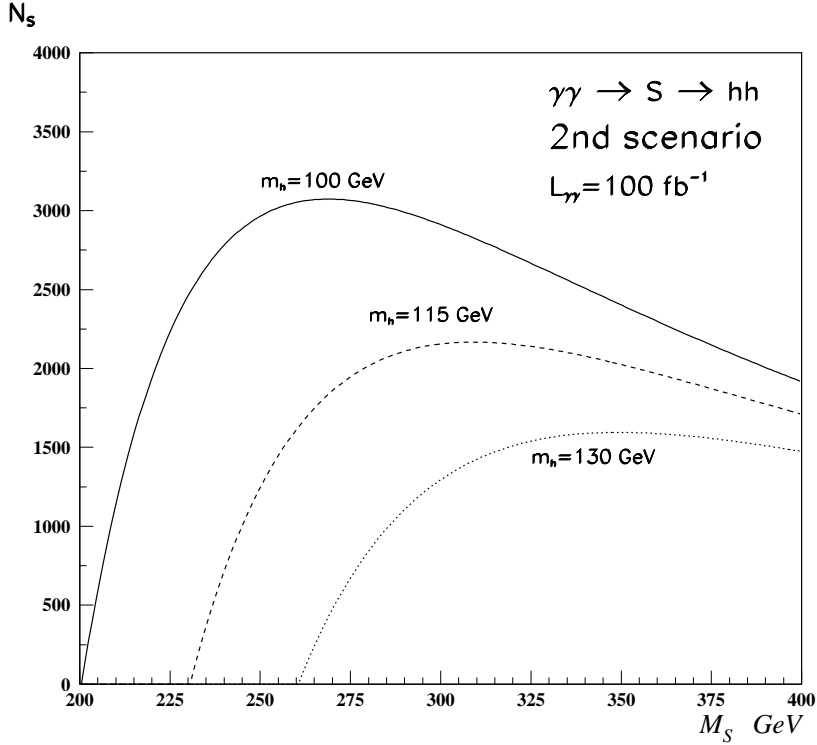,width=12cm}}
\end{center}
\caption{Total number of stoponium events in dominant $hh$-channel for
the second scenario.
Background (direct production of double light Higgses and $4b-jets$)
is expected less than 100 events.
}
\label{hh-fig}
\end{figure}

\vspace{0.2cm}  
\noindent  
{\bf 2.}  
In the second scenario stoponium total width could be about 10 MeV or even
larger. The photon-photon branching in this case is smaller, $\sim (2-4)\cdot
10^{-4}$. So, thousands of stoponiums will be produced per year and almost all
of them will decay to pairs of lightest Higgs bosons. We plot total number of
stoponium events at various $M_S$ and $m_h$ in Figure~\ref{hh-fig}. 

This result means, that stoponium will be discovered practically immediately
after PLC start, since the background ($hh$, $bbbb$ ...) is very small.

Note that in this scenario several years of operating at high luminosity is
needed in order to observe stoponium at LHC \cite{drees}.

Other branchings are at the same level as in the first scenario, or even
smaller. Thus, one can not hope to distinguish stoponium events in these
channels with statistics 100 fb$^{-1}$. Nevertheless, once stoponium has been
discovered the $gg$, $\gamma\gamma$ and $ZZ$ channels will exhibit
possibilities of study stoponium effective couplings with higher statistics.

\vspace{0.2cm} 
There is a range of parameters in the model with light stop quarks and large
$A_t$, where coupling between the lightest stops and the lightest Higgs boson
tends to increase in comparison with $t\bar{t}h$ in SM \cite{tth}, and which
correlates with the parameter range of scenario~2. So, one can expect some
enhancement of $hh$ production at PLC in this case due to stop-loop
contribution. However, in the  second scenario this stop-Higgs enhancement for
sure can not reduce excellent discovery potential of the dominant $hh$ decay
channel.

\vspace{1.5cm}
\noindent
{\large \bf Conclusions and further comments}
\vspace{0.2cm}

\noindent  
{\bf 1.}  
Main conclusion is that for some scenarios of supersymmetric extension of the
Standard Model photon linear collider will be the best machine to discover and
study bound state of stops if it exists.

In the decay channel into two lightest Higgs bosons (if it is permitted
kinematically) stoponiums will be observed at the beginning of PLC operating in
the case of dominant decay channel being $hh$, and during the first year if
$gg$ channel dominates.

In scenario with $hh$ decay mode being principal the stoponium discovery mass
range will be limited only by attainable values of the $\gamma\gamma$ collision
energy, which is discussed up to $0.8\cdot 500$ GeV. In this case PLC will be
a {\it stoponium factory} producing thousands of these heavy strong resonances
practically without background. Fine structure of their spectroscopy would be
available for study.

The tuning of PLC collision energy at the resonance point is necessary within
the 15\% window.

If $S\to hh$ mode is closed, the search for stoponiums would be not easy, this
bound state could be discovered in $gg$, $\gamma\gamma$ and $ZZ$
channels after a year. 
Anyway, if 500 fb$^{-1}$ of integrated statistics will be
really available for the first PLC these channels will allow to observe
stoponium for sure. 

All these channels ($hh$, $gg$, $\gamma\gamma$ and $ZZ$) have good chances for
probing the corresponding effective coupling constants  with statistics 500
fb$^{-1}$.

\vspace{0.2cm}  
\noindent  
{\bf 2.}  
A few comments can be made. The first one is related to the circumstance that
ground state of stoponium could not be distinguished from excited states due to
the detector resolutions. Therefore, the resonance peak will include
contributions from ground state and all excitations, leading to enhancement
factor of about 2~\cite{drees} in all cross sections. At the same time there
is a big uncertainty because of  poor understanding of the stoponium wave
function, that results in 30-50\% error when the stoponium rates are
estimated~\cite{drees}.

The second comment concerns the models with stoponium and the heaviest CP-even
Higgs boson masses being almost degenerate, $M_S\simeq m_H$. One can note that
there is no motivation for this scenario. Nevertheless, this degeneration
exhibits specific physics. In this case, all ``standard'' decay modes
(determined by the same mechanisms as for quarkoniums) are highly suppressed,
by a factor of $10^{-3}-10^{-4}$, due to huge increasing of the stoponium total
width saturated there by Higgs-originated channels, e.g. $b\bar b$ at high
$\tan\beta$ or $t\bar{t}$ at large $M_S$. Certainly, if the mass degeneration
is exact, $M_S=m_H$, quasistationary description of stoponium formation is
spoilt by too fast annihilation through Higgs boson into $b\bar{b}$
($t\bar{t}$). Nevertheless, in some vicinity of the degeneration point partial
width of the corresponding dominant decay channel could be enhanced not so
greatly and stoponium could be composed. In Eq.~(\ref{gamgam}) the
corresponding suppressing factor in the $\gamma\gamma$ branching will be
compensated by the same factor inverted in the stoponium total width. As a
result the signal rate in the main (Higgs originated) decay channel will be at
the level of the total stoponium production rate in the nondegenerate case:
140-50 fb for $M_S=200-300$ GeV. One can check that $\gamma\gamma\to b\bar b$
background has cross section about 1.8-0.9 pb for $\sqrt{s}=200-300$ GeV.
Therefore, stoponiums will be discovered in $b\bar b$ decay mode 
when PLC starts. At large $M_S$ in models with low and intermediate
$\tan\beta$, $t\bar{t}$ channel will dominate, and the stoponium production
will be about 10 fb for $M_S=A_t=500$~GeV. The background ($\gamma\gamma\to
t\bar t$) cross section is about 1pb, that yields the signal significance of 3
for statistics 100 fb$^{-1}$. Hence, in such model heavy stoponium will be
observed after accumulating of three years statistics. 

The third comment is dedicated to the fact that stoponium and CP-even Higgs
bosons have the same quantum numbers. Hence, first, they can be mixed  and
interfere in the collision experiments. The corresponding off-diagonal
insertion $\delta M^2$ into mass squared matrix may be estimated as
$A_tM_S(|\psi(0)|/M_S^{3/2})$, where $\psi(0)$ is stoponium wave function at
the origin. By making use of results of Ref.~\cite{ng}, where $\psi(0)$ was
evaluated, we obtain $\delta M^2\sim 2\% M_S A_t$ for $M_S\simeq 200-600$~GeV.
At $A_t=M_S=200$~GeV this contribution alters mass of the heaviest
CP-even Higgs
boson on 2~MeV-2~GeV depending on the degeneracy between stoponium and Higgs
masses (we consider $M_H<600$~GeV). Moreover, stoponium could be mixed with the
lightest Higgs boson leading to the contribution of a few MeV to the Higgs
mass. Note, that in LHC and linear $e^+e^-500$ experiments the mass of Higgs
boson will be measured with accuracy of about 1 GeV, or even better. Thus, the
effect of $M_H$ change due to {\it stoponium-Higgs} mixing could be valuable.
We stress, that this effect can be detected even if stoponium is not observed
directly. 

It is worth to note, that construction of muon collider (which is under
discussion now \cite{muoncoll}) with unique energy resolution will allow the
observation of the Higgs mass shifting at the level of tens MeV's (maybe even a
few MeV's). Consequently, if stop bound states exist, the sector of neutral
scalar resonances in supersymmetric models could have more complicated
structure than it could be deduced from the MSSM Lagrangian. This mixing effect
should be taken into account in the reconstruction of the Higgs potential
(parameters of the symmetry breaking mechanism).

Then, stoponium may be misidentified with Higgs boson, because there is a range
of parameters, where (dominant) partial widths of these two heavy objects 
almost coincide. In this case the observation of subdominant modes will be
required in order to distinguish these two resonances. 

\vspace{0.2mm}

The authors are indebted to V.~Kuzmin, A.~Penin and G.~Pivovarov for useful
discussions. The work of D.G. was supported in part under RFBR  grant
99-01-18410 and by the Russian Academy of Science, JRP grant \# 37. The work of
V.I. has been partially supported by the INTAS CERN99-0377 and RFBR-DFG
99-02-04011 grants and by St.Petersburg Grant Center.

\def\ijmp#1#2#3{{\it Int. Jour. Mod. Phys. }{\bf #1~} (19#2) #3}
\def\pl#1#2#3{{\it Phys. Lett. }{\bf B#1~} (19#2) #3}
\def\zp#1#2#3{{\it Z. Phys. }{\bf C#1~} (19#2) #3}
\def\prl#1#2#3{{\it Phys. Rev. Lett. }{\bf #1~} (19#2) #3}
\def\rmp#1#2#3{{\it Rev. Mod. Phys. }{\bf #1~} (19#2) #3}
\def\prep#1#2#3{{\it Phys. Rep. }{\bf #1~} (19#2) #3}
\def\pr#1#2#3{{\it Phys. Rev. }{\bf D#1~} (19#2) #3}
\def\np#1#2#3{{\it Nucl. Phys. }{\bf B#1~} (19#2) #3}
\def\mpl#1#2#3{{\it Mod. Phys. Lett. }{\bf #1~} (19#2) #3}
\def\arnps#1#2#3{{\it Annu. Rev. Nucl. Part. Sci. }{\bf #1~} (19#2) #3}
\def\sjnp#1#2#3{{\it Sov. J. Nucl. Phys. }{\bf #1~} (19#2) #3}
\def\jetp#1#2#3{{\it JETP Lett. }{\bf #1~} (19#2) #3}
\def\app#1#2#3{{\it Acta Phys. Polon. }{\bf #1~} (19#2) #3}
\def\rnc#1#2#3{{\it Riv. Nuovo Cim. }{\bf #1~} (19#2) #3}
\def\ap#1#2#3{{\it Ann. Phys. }{\bf #1~} (19#2) #3}
\def\ptp#1#2#3{{\it Prog. Theor. Phys. }{\bf #1~} (19#2) #3}
\def\spu#1#2#3{{\it Sov. Phys. Usp.}{\bf #1~} (19#2) #3}
\def\apj#1#2#3{{\it Ap. J.}{\bf #1~} (19#2) #3}
\def\epj#1#2#3{{\it Eur.\ Phys.\ J. }{\bf C#1~} (19#2) #3}
\def\pu#1#2#3{{\it Phys.-Usp. }{\bf #1~} (19#2) #3}

\end{document}